\documentclass[11pt]{article}
\usepackage{moriond,epsfig}

\input paperdef

\begin{document}
\vspace*{4cm}
\title{SUSY HIGGS BOSONS AT THE LHC}

\author{G. WEIGLEIN}

\address{IPPP, Department of Physics, University of Durham,\\
South Road, Durham DH1~3LE, UK}

\maketitle\abstracts{
Recent results on MSSM Higgs physics at the LHC are reviewed.
The dependence of the LHC discovery reach in the 
$b\bar b H/A, H/A \to \tau^+\tau^-$ channel
on the underlying SUSY
scenario is analysed. This is done by combining the latest results for 
the prospective CMS experimental sensitivities for an integrated
luminosity of 30 or 60~\ifb with state-of-the-art theoretical
predictions of MSSM Higgs-boson properties. The results are interpreted
in terms of the parameters governing the MSSM Higgs sector at lowest
order, $\MA$ and $\tb$. While the higgsino mass parameter $\mu$ has a
significant impact on the prospective discovery reach (and
correspondingly the ``LHC wedge'' region), it is found that the
discovery reach is rather stable with respect to variations
of other supersymmetric parameters.
Within the discovery region a determination of the masses of the 
heavy neutral Higgs bosons with an accuracy of 1--4\% seems feasible.
It is furthermore shown that Higgs-boson production in central exclusive 
diffractive channels can provide important information on the
properties of the neutral MSSM Higgs bosons.
}

\section{Introduction}

Signatures of an extended Higgs sector would provide unique evidence for
physics beyond the Standard Model (SM). While models with an extended Higgs
sector often give rise to a relatively light SM-like Higgs boson over a large
part of their parameter space, detecting heavy states of an extended
Higgs sector and studying their properties will be of utmost importance 
for revealing the underlying physics. 
%In the following, recent results
%on the physics of heavy neutral Higgs bosons of the Minimal
%Supersymmetric extension of the Standard Model (MSSM) at the LHC are
%briefly reviewed.

\section{Dependence of the LHC discovery reach on the SUSY scenario}

In \citere{higgscms} the reach of the CMS experiment with 30 or 60~\ifb\
for the heavy neutral MSSM Higgs bosons has been analysed 
focusing on the channel $b\bar b H/A, H/A \to \tau^+\tau^-$ with the
$\tau$'s subsequently decaying to jets and/or leptons. 
The experimental analysis, yielding the number of events
needed for a 5$\,\si$ discovery (depending on the mass of the Higgs
boson) was performed with full CMS detector
simulation and reconstruction for the final states of di-$\tau$-lepton
decays~\cite{CMSTDR}.
The events for the signal and background processes were generated using
PYTHIA~\cite{PYTHIA}.
%(except for the $\tau^{+} \tau^{-} b \bar b$
%background, which was generated using CompHEP~\cite{Boos:2004kh}).
The experimental analysis has been combined with predictions for the
Higgs-boson masses, production processes and decay channels obtained
with the code {\tt FeynHiggs}~\cite{feynhiggs}, taking into account all 
relevant higher-order corrections as well as possible decays of the heavy 
Higgs bosons into supersymmetric particles. 
The results
have been interpreted in terms of the two parameters $\tb$, the ratio of
the vacuum expectation values of the two Higgs doublets of the MSSM, and
$\MA$, the mass of the $\cp$-odd Higgs boson. The variation of the
discovery contours in the $\MA$--$\tb$ plane indicates the
dependence of the ``LHC wedge'' region, i.e.\
the region in which only the light $\cp$-even  MSSM Higgs boson can be
detected at the LHC at the 5$\,\si$ level, on the details of the
supersymmetric theory. 
See \citere{previous} for previous analyses. 

%%%%%%%%%%%%%%%%%% F I G U R E %%%%%%%%%%%%%%%%%%%%%%%%%%%%%%%%%%%%%%%%%%%%%%%
\begin{figure}[htb!]
\BC
\includegraphics[width=.35\textwidth]{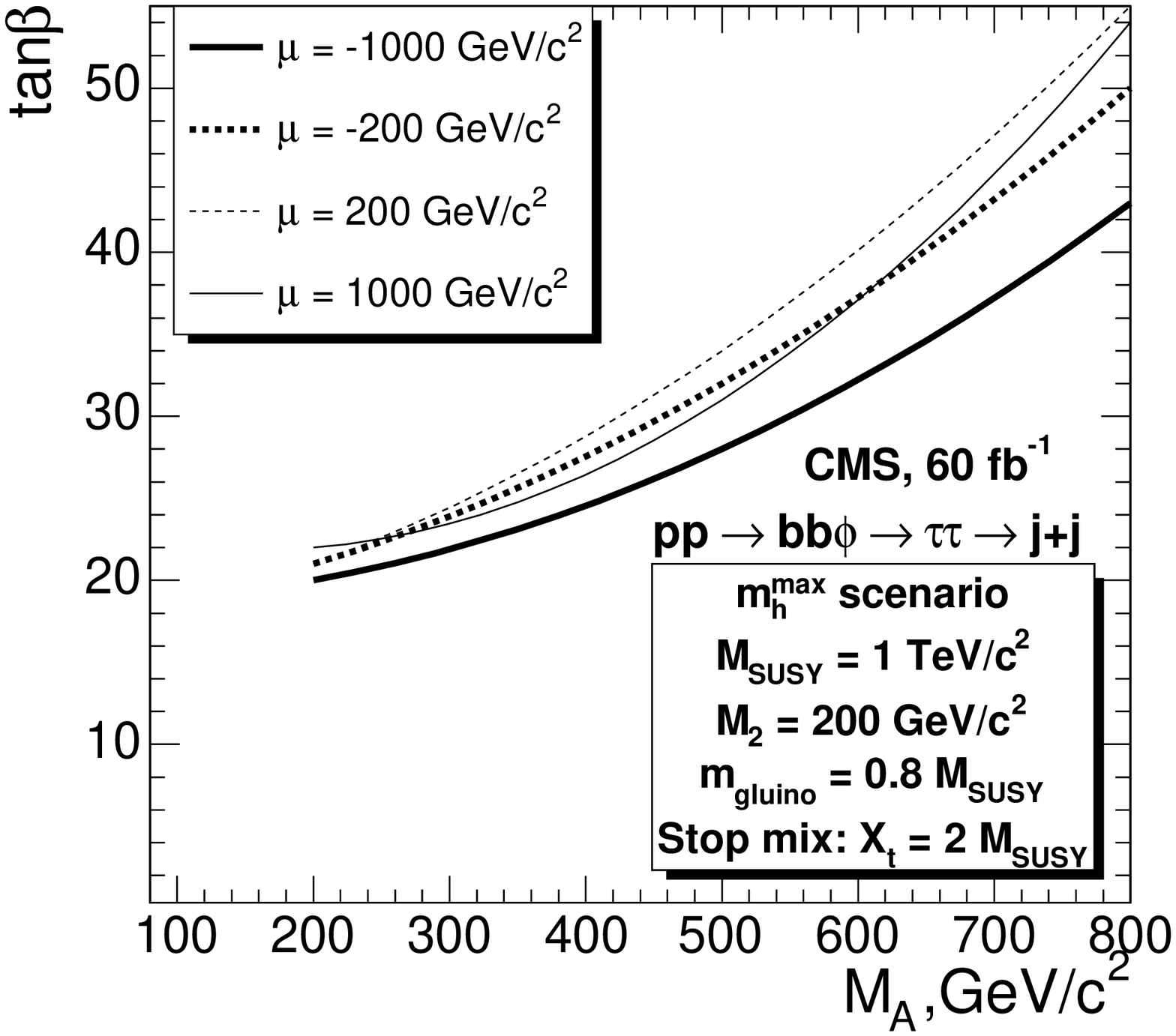}
\hspace{1em}
\includegraphics[width=.35\textwidth]{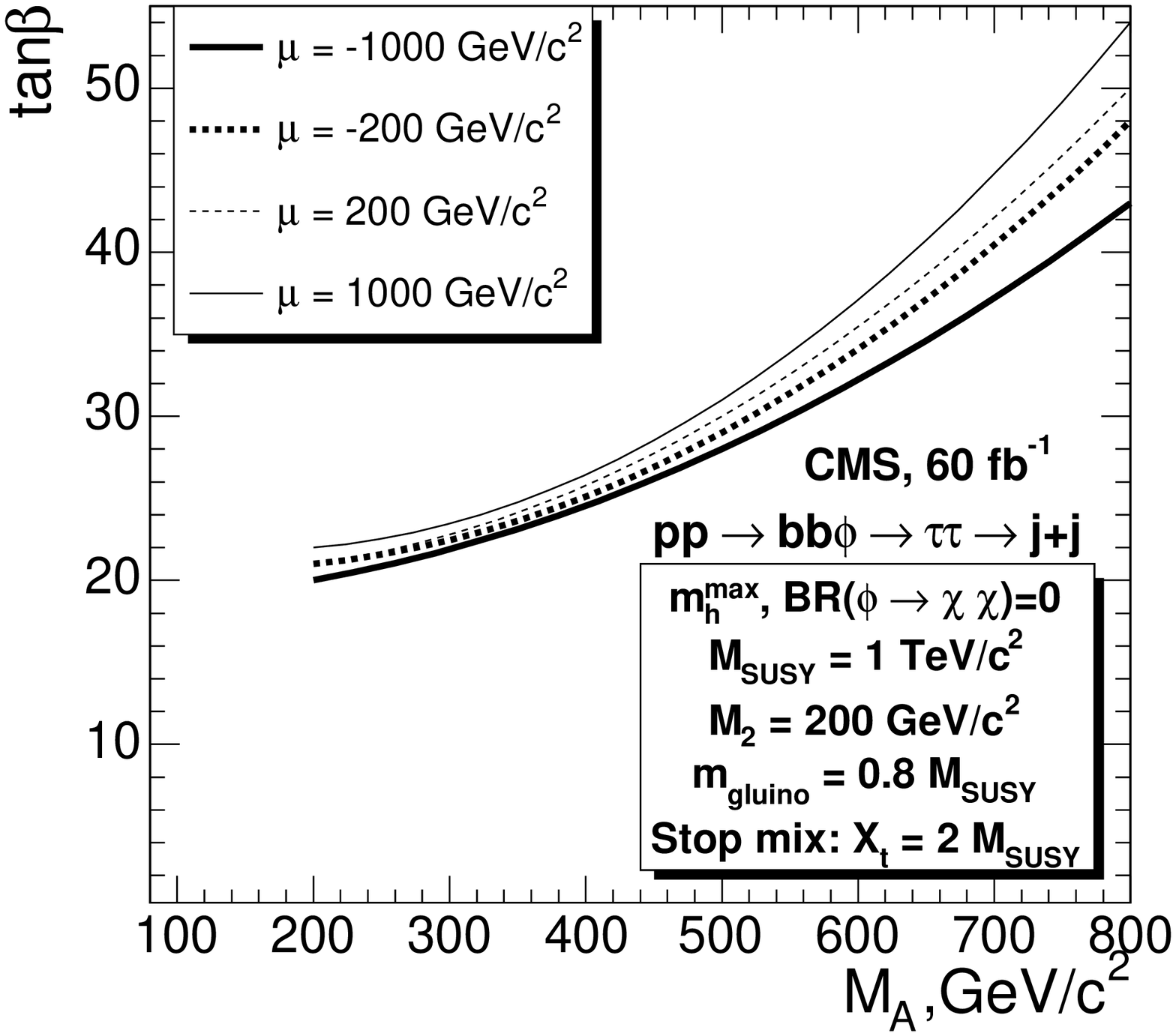}
\EC
\caption{Variation of the $5 \si$ discovery contours obtained
from the channel $b\bar b \phi, \phi \to \tau^+\tau^- \to \,\mbox{jets}$
in the $\mhmax$ benchmark scenario
for different values of~$\mu$ (left plot). The right plot shows the
result in the case where no decays of the heavy
Higgs bosons into supersymmetric particles are taken into account.
}
\label{fig:jj}
\end{figure}
%%%%%%%%%%%%%%%%%% F I G U R E %%%%%%%%%%%%%%%%%%%%%%%%%%%%%%%%%%%%%%%%%%%%%%%

\reffi{fig:jj} shows the variation of the $5 \si$ discovery contours
obtained
from the channel $b\bar b \phi, \phi \to \tau^+\tau^- \to \,\mbox{jets}$
in the 
$\mhmax$ benchmark scenario~\cite{benchmark2} for various values of the
higgsino mass parameter $\mu$. The parameter $\mu$ enters via
higher-order corrections affecting in particular the bottom Yukawa
coupling as well as via its kinematic effect in Higgs decays into
charginos and neutralinos. Both effects can be seen in \reffi{fig:jj}.
While the left plot shows the full result, in the right plot no decays
of the %heavy
Higgs bosons into supersymmetric particles are taken into account, so
that the right plot purely displays the effect of higher-order
corrections. Comparison of the two plots shows that in the region of
large $\tb$ (corresponding to larger values of $\MA$ on the discovery
contours) the dominant effect arises from higher-order corrections.
For lower values of $\tb$, on the other hand, the modification of the
Higgs branching ratio as a consequence of decays into supersymmetric
particles yields the dominant effect on the 5$\,\si$ discovery contours.
The largest shift in the 5$\,\si$ discovery contours amounts up to about
$\De\tb = 10$. The discovery contours have been shown to be rather
stable with respect to the impact of other supersymmetric 
contributions~\cite{higgscms}.

%%%%%%%%%%%%%%%%%% F I G U R E %%%%%%%%%%%%%%%%%%%%%%%%%%%%%%%%%%%%%%%%%%%%%%%
\begin{figure}[htb!]
\BC
\includegraphics[width=.35\textwidth]{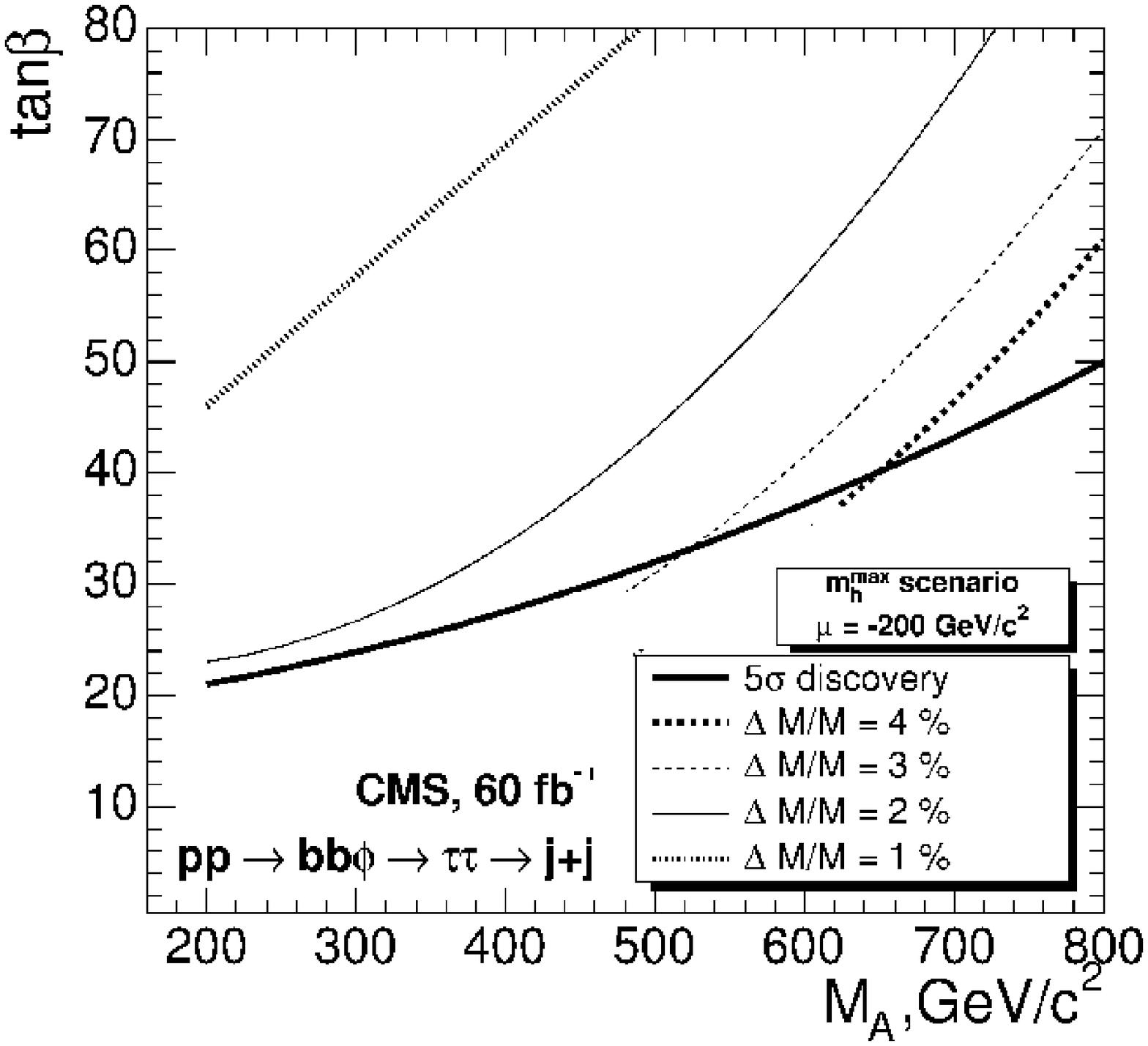}
\hspace{1em}
\includegraphics[width=.35\textwidth]{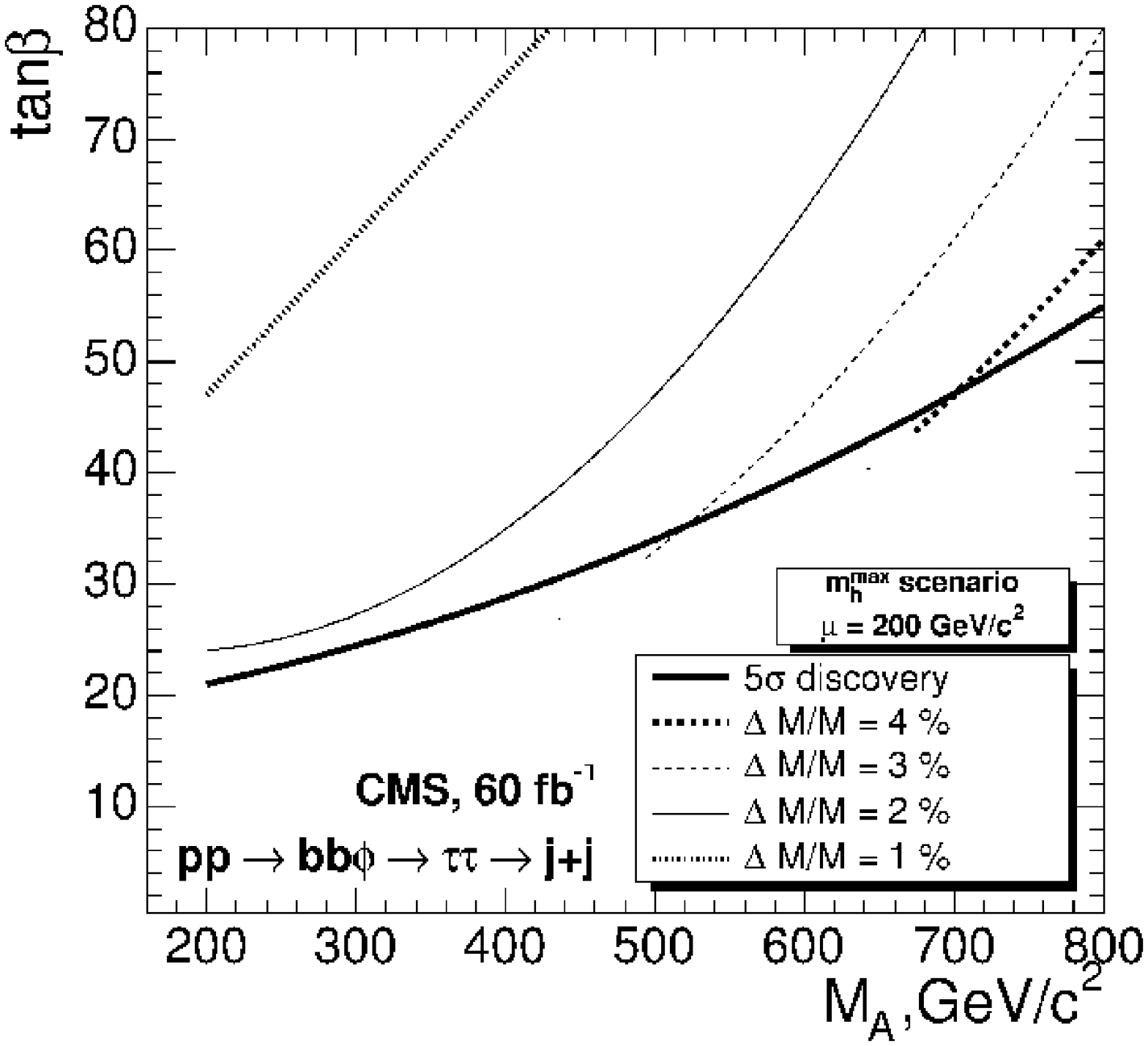}
\EC
\caption{The statistical precision of the Higgs-boson mass measurement
achievable
from the channel $b\bar b \phi, \phi \to \tau^+\tau^- \to \,\mbox{jets}$
in the $\mhmax$ benchmark scenario for $\mu = -200 \gev$ (left) and
$\mu = +200 \gev$ (right) is shown together with the 5$\,\si$ discovery
contour.}
\label{fig:jjmass}
\end{figure}

%%%%%%%%%%%%%%%%%% F I G U R E %%%%%%%%%%%%%%%%%%%%%%%%%%%%%%%%%%%%%%%%%%%%%%%

The prospective accuracy of the mass measurement of the
heavy neutral MSSM Higgs bosons in the channel
$b\bar b H/A, H/A \to \tau^+\tau^-$ is analysed in \reffi{fig:jjmass}.
The statistical accuracy of the mass measurement has been evaluated
via $\frac{\De M_\phi}{M_\phi} = \frac{R_{M_\phi}}{\sqrt{N_S}}$, where
$R_{M_\phi}$ is the ratio of the di-$\tau$ mass resolution to the 
Higgs-boson mass, and $N_S$ is the number of signal events 
($\phi = H, A$). \reffi{fig:jjmass} shows that statistical
experimental precisions of 1--4\% are
reachable within the discovery region. These results
are not expected to
considerably degrade if further uncertainties related to background
effects and jet and missing $E_{\rm T}$ scales are taken into account.
As discussed in \citere{higgscms}, a \%-level precision of the mass 
measurements could in favourable regions of the MSSM parameter allow to
experimentally resolve the signals of the two heavy MSSM Higgs bosons.

\section{MSSM Higgs bosons in central exclusive diffractive production}

Adding forward proton detectors to the CMS and ATLAS experiments 
(at 220~m and 420~m distance around them) can
complement the standard LHC physics menu in an interesting way. In
particular, ``central exclusive diffractive''
(CED) Higgs-boson production, where the outgoing protons remain
intact and there is no hadronic activity between them, profits from an
angular momentum selection rule~\cite{diffract1}
that permits a clean determination of
the quantum numbers of the
observed Higgs resonance which will be dominantly produced in a
scalar state. Other important features of the CED Higgs-boson production
process are a potentially excellent mass resolution (irrespective of 
the decay mode of the produced particle), a much better
signal-to-background ratio than conventional Higgs search channels at
the LHC, and the possibility to simultaneously access the $b \bar b$,
$WW$  and $\tau\tau$ decay modes of the Higgs boson(s).

%%%%%%%%%%%%%%%%%%%%%%%%%%%%%%%% Begin FIGURE %%%%%%%%%%%%%%%%%%%%%%%%%%%%%%%%%%
\begin{figure}[htb!]
\begin{center}
\includegraphics[width=9cm %,height=8.9cm
                ]{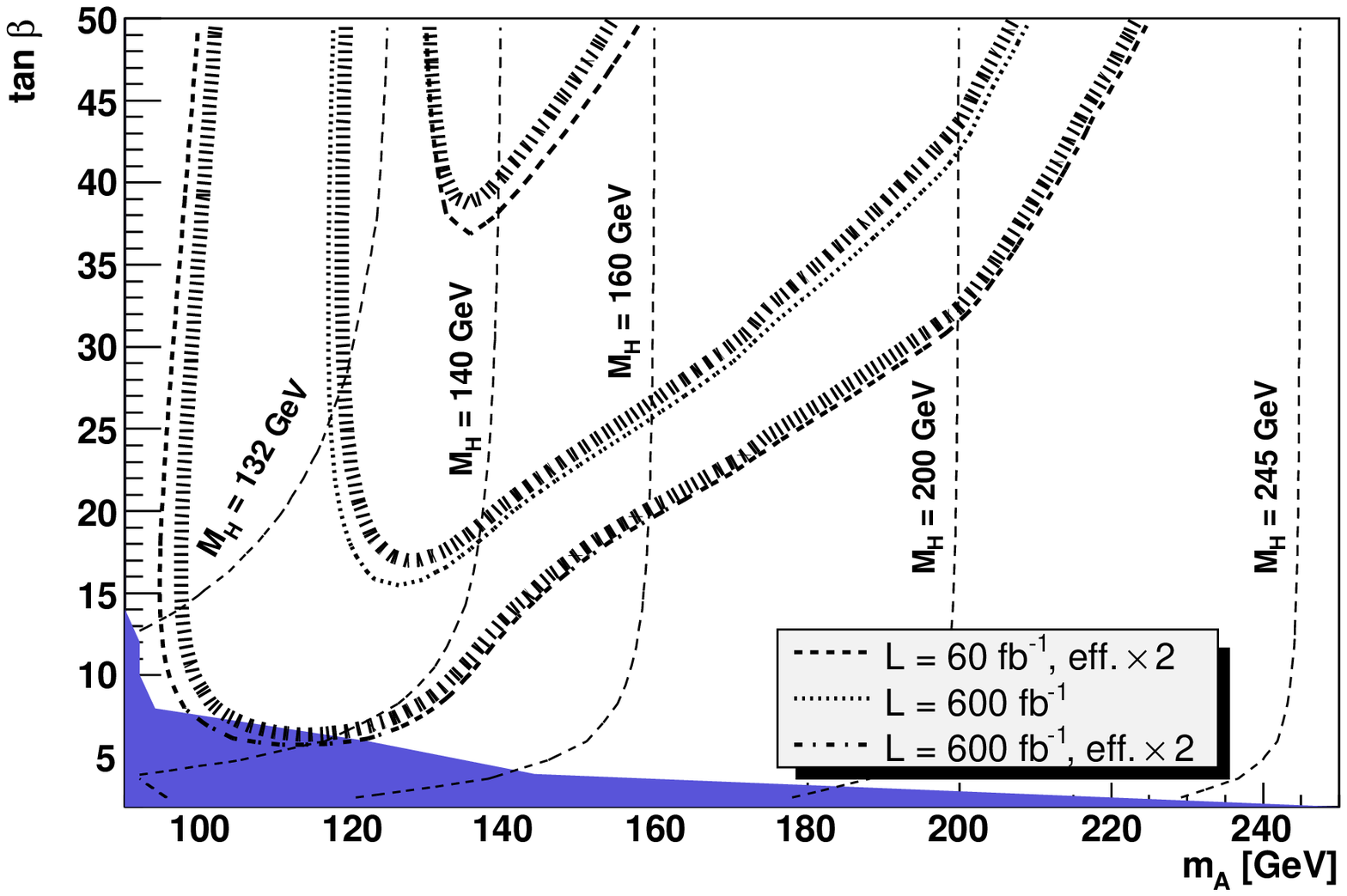}
\caption{
$5 \si$ discovery contours for the $H \to b \bar b$ channel in
CED production in the $\MA$--$\tb$ plane of the MSSM. The prospective
discovery reach is
shown in the $\mhmax$ benchmark scenario (with $\mu = +200 \gev$),
The results are shown for assumed effective luminosities (combining
ATLAS and CMS) of 60~\ifb,
60~\ifb$\times 2$, 600~\ifb and 600~\ifb$\times 2$.
The values of the mass of the heavy $\cp$-even Higgs boson, $\MH$, are
indicated by contour lines. The dark shaded (blue)
region corresponds to the parameter region that
is excluded by the LEP Higgs searches in the channel
$e^+e^- \to Z^* \to Z h, H$.
}
\label{fig:Hbb200}
\end{center}
\end{figure}
%%%%%%%%%%%%%%%%%%%%%%%%%%%%%%%% End FIGURE %%%%%%%%%%%%%%%%%%%%%%%%%%%%%%%%%%%%

In \citere{higgsdiffract} a detailed investigation of the 
prospects for the MSSM Higgs-boson channels
$h, H \to b \bar b, \tau^+\tau^-, W W^*$ in CED production
has been carried out (for other studies in the MSSM, see
\citere{diffract2} and references therein). 
In CED the heavy $\cp$-even MSSM Higgs boson $H$ can be produced and its
decay into $b \bar b$ can be utilised. While in the SM the
${\rm BR}(H \to b \bar b)$ is strongly suppressed for $\MH \gsim 2 \MW$
because of the dominant decay into gauge bosons, in the MSSM
$H \to b \bar b$ remains by far the dominant decay mode also for larger
masses as long as no decays into supersymmetric particles (or lighter
Higgs bosons) are open. The CED Higgs-boson production in the $b \bar b$
channel is therefore important over a much larger mass range than in the
SM. As an example,
\reffi{fig:Hbb200} shows the 
$5 \si$ discovery contours for the $H \to b \bar b$ channel in
CED production in the $\MA$--$\tb$ plane of the MSSM (using the 
$\mhmax$ benchmark scenario~\cite{benchmark2}) for different luminosity
scenarios. It is found that the CED Higgs-boson production channel can
cover an interesting part of the MSSM parameter space at the $5 \si$
level if the CED channel can be utilised at high instantaneous
luminosity (which requires in particular to bring pile-up background
under control). 
For an effective luminosity of 600~\ifb$\times 2$ (see
\citere{higgsdiffract}) the
discovery of a heavy $\cp$-even
Higgs boson with a mass of about $140 \gev$ will be possible for all
values of $\tb$. This is of particular interest in view of the ``wedge
region'' left uncovered by the conventional search channels for heavy
MSSM Higgs bosons (see above).
In the high-$\tb$ region the discovery reach extends beyond 
$\MH = 200\gev$ at the 5-$\si$ level. If the Higgs bosons $h$ and/or
$H$ have been detected in the conventional search channels, a lower
statistical significance may be sufficient for the CED production of
$h$ and $H$, corresponding to a larger coverage in
the $\MA$--$\tb$ plane. The CED Higgs-boson production channel will
provide in this case important information on the Higgs-boson properties 
and may even allow a direct measurement of the Higgs-boson 
width~\cite{higgsdiffract}.

\subsection*{Acknowledgments}
The author gratefully acknowledges the collaboration with 
S.~Gennai, S.~Heinemeyer, A.~Kalinowski, 
V.A.~Khoze, 
R.~Kinnunen, S.~Lehti, A.~Nikitenko,
M.G.~Ryskin, W.J.~Stirling and M.~Tasevsky
on the results presented in this paper.
He also thanks the organisers of the 42nd Rencontres de Moriond
for the kind invitation and the pleasant atmosphere at the meeting.

\end{document}

%%%%%%%%%%%%%%%%%%%%%%
% End of moriond.tex  %
%%%%%%%%%%%%%%%%%%%%%%